\documentclass[prl,floatfix,amsmath,amsfonts,twocolumn,showpacs]{revtex4}
\usepackage{dcolumn} 
\usepackage{bm}
\usepackage{epsfig}
\newcommand{\beq}{\begin{equation}}
\newcommand{\eeq}{\end{equation}}
\newcommand{\beqa}{\begin{eqnarray}}
\newcommand{\eeqa}{\end{eqnarray}}
\newcommand{\ba}{\begin{array}}
\newcommand{\ea}{\end{array}}
\usepackage{color}
\begin{document}
\title{Matter-wave 2D solitons in crossed linear and nonlinear optical lattices
}
\author{H.L.F. da Luz$^{1}$}
\author{F.Kh. Abdullaev$^{2}$}
\author{A. Gammal$^{1}$}
\author{M. Salerno$^{3}$}
\author{Lauro Tomio $^{4,5}$}
\affiliation{
\\
$^1$ Instituto de F\'isica, Universidade de S\~ao Paulo, 05508-090,
S\~ao Paulo, SP, Brazil.\\
$^2$CFTC, Complexo Interdisciplinar, Universidade Lisboa, Av.Prof. Gama Pinto, 2, 1649-003
Lisboa, Portugal.\\
$^3$Dipartimento di Fisica ``E.R. Caianiello", CNISM  and INFN -
Gruppo Collegato di Salerno, Universit\`a di Salerno, Via Ponte
don Melillo, 84084 Fisciano (SA), Italy.\\
$^4$Instituto de F\'{\i}sica, Universidade Federal
Fluminense, 24210-346, Niter\'oi, RJ, Brazil.\\
$^5$Instituto de F\'\i sica Te\'orica, Universidade Estadual Paulista
(UNESP), 01140-070, S\~ao Paulo, SP, Brazil.
}
\begin{abstract}
It is demonstrated the existence of multidimensional matter-wave solitons in
a crossed optical lattice (OL) with linear OL in the $x-$direction and
nonlinear OL (NOL) in the $y-$direction, where the NOL can be generated by a
periodic spatial modulation of the scattering length using an optically induced
Feshbach resonance.
In particular, we show that such crossed linear and nonlinear OL allows to stabilize
two-dimensional (2D) solitons against decay or collapse for both  attractive and
repulsive interactions.
The solutions for the soliton stability are investigated analytically,
by using a multi-Gaussian variational approach (VA), with the Vakhitov-Kolokolov (VK)
necessary criterion for stability; and numerically, by using the relaxation method and
direct numerical time integrations of the Gross-Pitaevskii equation (GPE).
Very good agreement of the results corresponding to both treatments is observed.
\end{abstract}
 \maketitle

\section{Introduction}

Bose-Einstein condensates (BEC) in optical lattices (OL) are presently attracting
a great deal of  interest, both experimentally and theoretically~\cite{Morsh,BK},
in connection with a series of physical phenomena which arise in condensed matter
physics, these including Bloch oscillations~\cite{BOLin,Pitaevskii,SKB}, generation
of coherent atomic pulses (atom laser)~\cite{Kasevich}, dynamical
localization~\cite{DL1-Arimondo,BKS-DL09,DL2-Arimondo}, Landau-Zener
tunneling~\cite{LZ-Arim,LZ-KKS05,LZ-Arimondo09}, superfluid Mott transitions~\cite{Gren},
etc. The flexibility of  BEC systems with respect to changes of parameters permits, 
indeed, to investigate these phenomena in much an easier way than in their condensed 
matter physics counterparts. On the other hand, BEC systems are intrinsically nonlinear 
and the correspondence with usual condensed matter phenomena can be established mainly
in the linear regime when the interatomic interactions are detuned to zero  by means
of external magnetic fields using the Feshbach resonance technique \cite{Inouye}.
The presence of nonlinearity, however, represents an additional resource for BEC
systems which leads to interesting  phenomena such as the existence of localized
bound states which remain stable for infinite time due to a perfect balance between
nonlinearity and dispersion. In presence of a periodic potential or a linear optical
lattice (LOL),  these states are  known as gap-solitons (they have chemical potentials located inside the band-gaps of the underlying linear band structure),
which can exist for both attractive and repulsive interactions~\cite{TS,ABDKS,Carus}, this last fact being  only possible due to the presence of the periodic potential. In this context, it has been shown that gap-solitons are formed through the mechanism of modulational instability of the Bloch states, at the edges of the Brillouin zone of the underlying linear periodic system~\cite{KS02}.
Their  existence in BECs with repulsive interactions was experimentally
demonstrated in Ref.~\cite{Eier}. Recently, the possibility of Bloch oscillations~\cite{SKB}, Rabi oscillations~\cite{BKS-PRA09} and dynamical localization~\cite{BKS-DL09} of BEC gap solitons, in presence of constant or time dependent linear forces (induced by accelerations of the OL), have also been considered.
In these cases, besides the LOL, periodic spatial variations of the nonlinearity,
also known as nonlinear optical lattice (NOL), have been used in order to avoid
dynamical instabilities and to make the above  phenomena long lived in the nonlinear regime.

In higher dimensions, LOLs were shown to be effective in stabilizing localized states against collapse or decay, leading to the formation of stable multidimensional
gap-solitons~\cite{BKS02,BMS03}. In particular, it has been shown that a one-dimensional
(1D) LOL in three-dimensional (3D) space does not allow to stabilize 3D gap-solitons. This soliton stabilization becomes only possible for attractive 3D BEC subjected to the action of a 2D LOL~\cite{BMS03,BMS04}. The possibility to stabilize solitons by means of NOL has been investigated mainly in the 1D case,  for which the mathematical properties have been studied in detail, for the ground state when considering solitons supported by 1D NOL~\cite{SM,AG,Fibich,Garcia}, as well as by combinations of linear and nonlinear OLs~\cite{abdul}.
In the 2D case, it was shown in \cite{AGLT07} that, for conservative systems, a NOL in one
direction by itself cannot give stable localized solutions in the case of attractive interactions. This fact remains  true also for 2D NOLs both for attractive and repulsive interactions.

Since NOLs can be created by means of external magnetic fields via the usual Feshbach resonance technique, or by optically induced Feshbach resonances (OFR)~\cite{OFR2}, it is of  interest to investigate whether crossed combined linear and nonlinear optical lattices (e.g.,
a LOL in one direction plus a NOL in the orthogonal direction) allow to stabilize multidimensional solitons. This analysis can be relevant in the perspective of experimental observations of
multidimensional solitons.  In this respect, we remark  that an experimental realization of a NOL  was recently achieved (on the submicron scale) for a $^{174}$Yb BEC  using the optical  Feshbach resonance technique~\cite{Yamazaki}. Moreover, we recall that to date no multidimensional BEC solitons in LOLs  have  been observed. Besides, giving the possibility of exploring alternate methods for multidimensional soliton creation, crossed combined linear and nonlinear OLs can allow to extend to a 2D setting  interesting  transmission and reflection phenomena of soliton wavepackets, such as  matter-wave optical limiting processes and bistability phenomena considered in the 1D case for potential applications as matter-wave limiters, BEC mirrors or cavities and atomic switches~\cite{dong07,fatkh-pre,He}.

The aim of the present paper is to investigate the existence of 2D matter-wave solitons in crossed OLs consisting  of a LOL in
the $x-$direction and  a NOL in the $y-$direction. In particular, we show  that
crossed linear and nonlinear OLs allow to stabilize 2D solitons both for
attractive and repulsive interatomic interactions.
This will be demonstrated by means of  analytical and numerical approaches: 
analytically, with the variational method and a multi-Gaussian ansatz; numerically, with
relaxation methods and  direct numerical integrations  of the Gross-Pitaevskii (GP) differential equation. Due to the lattice anisotropy, the solitons display elliptical cross
sections, which in our variational approach (VA) are accounted by a
multi-Gaussian ansatz with different parameters for each spatial directions.
Existent VA curves and corresponding stability properties of multidimensional solitons are given for several parameter values  in terms of chemical potentials and total
energies as functions of the number of  particles, using  the Vakhitov-Kolokolov (VK) criterion for the stability. The  results are then compared with direct numerical integrations of the full GP equation.

The paper is structured as follows. In section II, we introduce the model equations and
discuss the physical implementation of a 2D crossed linear and nonlinear lattice by using spatial modulations of the scattering length. In Section III,  we use a multi-Gaussian variational approach to derive our results, presented for chemical potentials and total energies as functions of the number of atoms, for the case of 2D solitons in 2D crossed OLs, for both attractive and repulsive interactions, where
the stability is investigated by the well known VK criterion. In Section IV, the
results of the VA  are compared with the ones obtained by direct integrations of the
GPE, using both relaxation in imaginary time and real time propagations to check
the stability. Finally, in Section V, the main results of the paper are
briefly summarized.

\section{Model equations}
Multidimensional BECs in 2D crossed linear and nonlinear OLs are described in the
mean field approximation by the following Gross-Pitaevskii equation (GPE):
\beq
i \hbar \frac{\partial \psi}{\partial t}= - \frac{\hbar^2}{2 m} \nabla_{\cal{D}}^2 \psi -
\Lambda\cos(2 k \,x) \psi - g(y)|\psi|^2 \psi
,\label{2DGPE0} \eeq
where $\psi\equiv\psi({\bf r},t)$ is normalized to the number of atoms,
$\nabla^2_{\cal D}$ denotes the Laplacian in dimension $\cal{D}=$ 2, $m$ is the
atomic mass, and  $\Lambda\cos(2k \,x)$ is a LOL in the $x-$direction, with
strength $\Lambda$ and lattice constant $\pi/k$. The function  $g(y)$ represents a NOL in
the $y-$direction, produced either by spatially varying magnetic fields near a
Feshbach resonance or by optically induced Feshbach resonances \cite{OFR2}.
It the following we assume $g(y)$ of the form
\beq
g(y)= g_0 + g_1 \cos(2 \kappa  y),
\label{NOL}
\eeq
where $g_0$ denotes the mean nonlinearity related to the mean $s-$wave scattering
length $a_{s0}$, and $g_1$ is the strength of a periodic modulation of the nonlinearity
in the $y-$direction, having the period $\pi /\kappa$.
 Note that the quasi-2D system
is confined in the $z-$direction, with the effective 2D scattering length being given
in terms of the 3D scattering length and some typical scale in the confined direction~\cite{Petrov2001,Lee2002}. In this case, the parameters $g_0$ and $g_1$ are
given in units of energy multiplied by some squared distance.
The spatial modulation can be produced by manipulating the scattering length with a laser
field  tuned near a photo association transition, e.g., close to the resonance of one of
the bound $p$ levels of the excited molecules. Virtual radiative
transitions of a pair of interacting atoms to this level can
change the value and even  reverse the sign of the scattering
length. One can show that a periodic variation of the laser field intensity applied
in the $y-$direction of the form $I(y)=I_0 \cos^2(\kappa y)$ produces a periodic
variation of the atomic scattering length, such that
$a_s(y) = a_{s0}[1 + \alpha I/(\delta +I)]$,
where $a_{s0}$ is the scattering length in the absence of light, $\delta$ is
the frequency detuning of the light from the resonance, and $\alpha$ is a
constant factor \cite{OFR2,SM}. For weak intensities, when $I_0 \ll |\delta|$, we have
that the real part of the scattering length can be approximated by
$a_s=a_{s0}+ a_{s1} \cos^2(\kappa y)$, leading to a modulated nonlinearity of
essentially the same form  assumed in Eq. (\ref{NOL}).

It should be remarked, however, that the creation of a NOL in a BEC, by manipulation
of the scattering length in this manner, also implies some spontaneous emission loss,
which is inherent in the optical Feshbach resonance technique\cite{Fatemi}-\cite{Thal}.
Such dissipative effects can be strongly reduced by using laser fields with sufficiently
high intensity and detuned from the resonance~\cite{Bauer}. It is also worth pointing out
that, by using a laser field to control a magnetic FR the losses can also be essentially
reduced in comparison with optically induced FR.
In particular, the experiment reported in \cite{Bauer} demonstrates that a laser light,
near the resonance with a molecular bound-to-bound transition in $^{87}$Rb, can be used
to shift the value of the magnetic field where the FR occurs. By this way, it is possible
to vary the scattering length on the optical wavelength scale, without having considerable
losses (about two orders of magnitude lower then in optically induced FR experiments).
We also remark that periodic variations of the scattering length, induced by the usual
Feshbach resonance technique using spatially periodic external magnetic fields, would
lead to similar conclusions.

In the following, we adopt dimensionless units by rescaling the space and time variables,
such that the variables in Eq.~(\ref{2DGPE0}) are redefined according to
$t \rightarrow (\hbar/E_r)\,t$,  ${\mathbf r} \rightarrow {\mathbf r}/k$, with
$E_r\equiv \hbar^2 k^2/ 2m$ being the recoil energy and ${\mathbf r}\equiv \{x,y,z\}$.
The wavefunction is rescaled as $\psi({\mathbf r},t) \rightarrow k u({\mathbf r},t)$, in terms of which the Eq.~(\ref{2DGPE0}) acquires the form
\beq
{\rm i}\frac{\partial u}{\partial t}= - \left[\frac{\partial^2 u}{\partial x^2} +\frac{\partial^2 u}{\partial y^2}\right]- V_l (x) u - \Gamma(y) |u|^2 u
,\label{2DGPE}
\eeq
where
 \beq
V_l(x)= \varepsilon \cos(2 x),\;\;\;
\Gamma(y)= \chi + \gamma \cos(\lambda\,  y),
\label{NOL2}
\eeq
denote the linear and nonlinear OL interactions, respectively.
In the above, $\varepsilon=\Lambda/E_r$,  $\chi= g_0 k^2/E_r = 8 \pi a_{s0} k$,
$\lambda=2 \kappa/k $,  and $\gamma=g_1 k^2/E_r$, with
$g_1$ assumed as a free parameter.

\section{Variational Analysis}
We consider localized solutions of Eq. (\ref{2DGPE}) in the  crossed linear and nonlinear OLs given by Eq. (\ref{NOL2}), in the cases that the mean nonlinearity
$\chi$ can be both negative (repulsive interaction) or positive (attractive interaction). In order to obtain analytical estimates for the soliton existence
and stability, we use  the variational approach (VA) and  look for solutions of
the form
\beq u(x,y,t)= U(x,y)\exp(-{\rm i}\mu t)\label{VAsol},\eeq
where $\mu$ is the chemical potential and $U(x,y)$ is a  real function for
the soliton profile.
Due to the anisotropy of the crossed OL, having different natures and different strengths in the two directions, we expect elliptical cross sections for the
soliton profiles. This feature can be accounted in our analysis of Eq.~(\ref{VAsol})
by adopting the following ansatz for $U(x,y)$:
\beq
U(x,y)= A e^{- (a x^2 + b y^2)/2}
,\label{ansatz}
\eeq
where $a$ and $b$ are the parameters controlling the Gaussian widths in the
two directions, with $a/b$ the aspect ratio of the solution. Using this ansatz,
the stationary GPE can be written as
{\small
\beq
\mu U + \frac{\partial^2 U}{\partial x^2}+ \frac{\partial^2 U}{\partial y^2}
+ \varepsilon \cos(2\, x) U + [\chi+ \gamma
\cos(\lambda \, y)] U^3 =0
,\label{statGPE}
\eeq}
from which the chemical potential and corresponding total energy are,
respectively, given by
{\small
\beq
\mu N =\int\left\{|\nabla U|^2  - \varepsilon \cos(2 x) U^2 - [\chi+ \gamma
\cos(\lambda\, y)] U^4\right\}dx\,dy ,\nonumber
\eeq
\beq
E=\int\left\{|\nabla U|^2  - \varepsilon \cos(2 x) U^2 -\frac{\chi+ \gamma
\cos(\lambda\, y)}{2}U^4\right\}dx\,dy .\nonumber
\eeq }

The integrations in the above and following expressions cover the 2D phase
space, from $-\infty$ to $+\infty$.
From the ansatz (\ref{ansatz}) it follows that the normalized number of atoms is
expressed in terms of the variational parameters $A, a, b$, as:
\beq
N = \int U^2 dx \, dy= \frac{A^2 \pi}{\sqrt{a b}}.
\eeq
The  stationary GPE in (\ref{statGPE}) can be derived  from the following field Lagrangian
\begin{eqnarray}
L &=\displaystyle\frac 12&\int\left\{|\nabla U|^2 - [\mu+ \varepsilon \cos(2 x)] U^2 \right.
\nonumber \\ && \left. -\frac 12 \left[\chi + \gamma \cos (\lambda \, y )\right] U^4\right\} dx \, dy.
\end{eqnarray}
By substituting the ansatz (\ref{ansatz}) into the above Lagrangian and performing
the integrations, in terms of the variational parameters $a$, $b$ and $A$, we obtain
the following effective Lagrangian:
\begin{eqnarray}
\label{leff}
L_{eff} &=\displaystyle\frac{A^2 \pi}{4 \sqrt{a b}} &\left\{a + b - \frac{A^2}{2}
\left[\chi + \gamma e^{-{\lambda^2}/{(8 b)}}\right] \right.
\nonumber \\
&&\left.  - 2\left[\mu + \varepsilon e^{-{1}/{a}}\right]\right\},
\end{eqnarray}
with the corresponding total energy given by
\begin{eqnarray}
E =N\left\{\frac{a + b}{2} - \frac{A^2}{4}
\left[\chi + \gamma e^{-{\lambda^2}/{(8 b)}}\right]-\varepsilon e^{-{1}/{a}}\right\}.
\end{eqnarray}
Then, following from $L_{eff}$, by using the Euler-Lagrange equations for the
parameters $a,b, A$, we can derive the relations for the chemical potential,
total energy and number of atoms. One can easily show that these relations can be
written as
\begin{eqnarray}
\mu  &=& \frac {b-3a}2 - \varepsilon \left(1- \frac {4}{a}\right) e^{-\frac{1}{a}} \label{va1},\\
N &=& \frac{4 \pi}{a \sqrt{a b}}\; \frac{ a^2-  2 \varepsilon  e^{-\frac{1}{a}} }
{\chi + \gamma  e^{-\frac{\lambda^2}{8 b}}}, \label{va2}\\
\frac{E}{N}&=&\frac {b-a}{2} - \varepsilon \left(1- \frac {2}{a}\right) e^{-\frac{1}{a}} \label{va3},
\end{eqnarray}
with $a, b$ solutions of the transcendental equation
\beq
\frac 1b + \gamma  \frac{\lambda^2}{4 b^2}\; \frac {e^{-\frac{\lambda^2}{8 b}}}
{\chi + \gamma  e^{-\frac{\lambda^2}{8 b}}}= F(a)
,\label{trascendent}
\eeq
where
\beq
F(a)=\frac{a}{a^2- 2 \varepsilon e^{-\frac{1}{a}}}.
\label{Fa}\eeq
Exact analytical solutions of Eq. (\ref{trascendent}) can be obtained
in the following limiting.

i) Case $\gamma=0$. In this case only the LOL in the x-direction is present and  from Eq. (\ref{trascendent}) we obtain $b=1/F(a)$. By
substituting this value of $b$ in Eqs. (\ref{va1}) and  (\ref{va2}), we get
the following parametric equations, in the $\mu-N$ space, for 
the existent curves of a soliton:
\begin{eqnarray}
\mu  &=& -a - \varepsilon \left(1- \frac {3 }{a}\right) e^{-\frac{1}{a}}, \label{vag01}\\
N &=& \frac{4 \pi}{\chi} \sqrt{1- \varepsilon \frac{2}{a^2} e^{-\frac{1}{a}}}. \label{vag02}
\end{eqnarray}
Notice from Eq. (\ref{leff}), that this case becomes  equivalent to the case 
with $\lambda=0$ and $\gamma\ne 0$, by replacing $\chi$ in (\ref{vag02}) 
with $\chi+\gamma$.
The above equations coincide with the ones of the 1D potential case investigated in
Ref.~\cite{BMS04} [compare the above equations with Eqs. (8) and (9) of this paper],
where it was shown that the system can support 2D localized solutions only for
attractive interactions. The absence of a confining potential in the $y-$direction
makes the solution to be uniform in this direction.  But, interestingly enough,
the solution is stable if the periodic boundary conditions are adopted in the
$y-$direction (line-soliton).

ii) Case $\chi=0$. In this case both the linear and  the nonlinear OL are present but the mean nonlinearity is detuned to zero. Then, the Eq.~(\ref{trascendent})
reduces to $4 b + \lambda^2= 4 b^2 F(a)$ and $b$ can be expressed in terms 
of the parameter $a$ as
\beq
b=\frac{1+\sqrt{1+ \lambda^2 F(a)}}{2 F(a)}.
\eeq

iii) Case $\varepsilon=0$. In this case only the NOL in the y-direction is present and  from Eq. (\ref{trascendent}) we have that  $F(a)=1/a$. The solutions for $\mu$ and $N$ can be conveniently expressed in terms of the parameter $b$ as:
\beq
\frac{a}{b}= \left[1+ \frac{\lambda^2}{4 b}
\left(1-
\; \frac {\chi}
{\chi + \gamma  e^{-\frac{\lambda^2}{8 b}}}\right)
\right]^{-1} ,
\label{e03}
\eeq
\begin{eqnarray}
\mu  &=& \frac {b}{2}\left(1-3\frac{a}{b}\right),  \label{e01}\\
N &=& {4 \pi}{\sqrt{\frac a b}}\; \frac{1}
{\chi + \gamma  e^{-\frac{\lambda^2}{8 b}}}. \label{e02}
\end{eqnarray}
In the absence of any confinement in the $x-$ and $y-$directions (e.g., for 
$\varepsilon=0$ and $\gamma=0$) the above equations reproduce the known results of the 2D nonlinear Schr\"odinger solitons.  In particular, for attractive interactions
($\chi>0$), the soliton widths in the two directions  become equal ($a=b$)  and the number of
atoms times the nonlinearity reduces to a constant, which coincides with the norm of 2D  Townes
solitons \cite{townes} determining the critical threshold for collapse  of a soliton in 2D as $N_{cr}=4 \pi/\chi$.

In the general case the solution of Eq. (\ref{trascendent}) must be found numerically.
\begin{figure}
\centerline{
\includegraphics[width=4.2cm,height=4.3cm,clip]{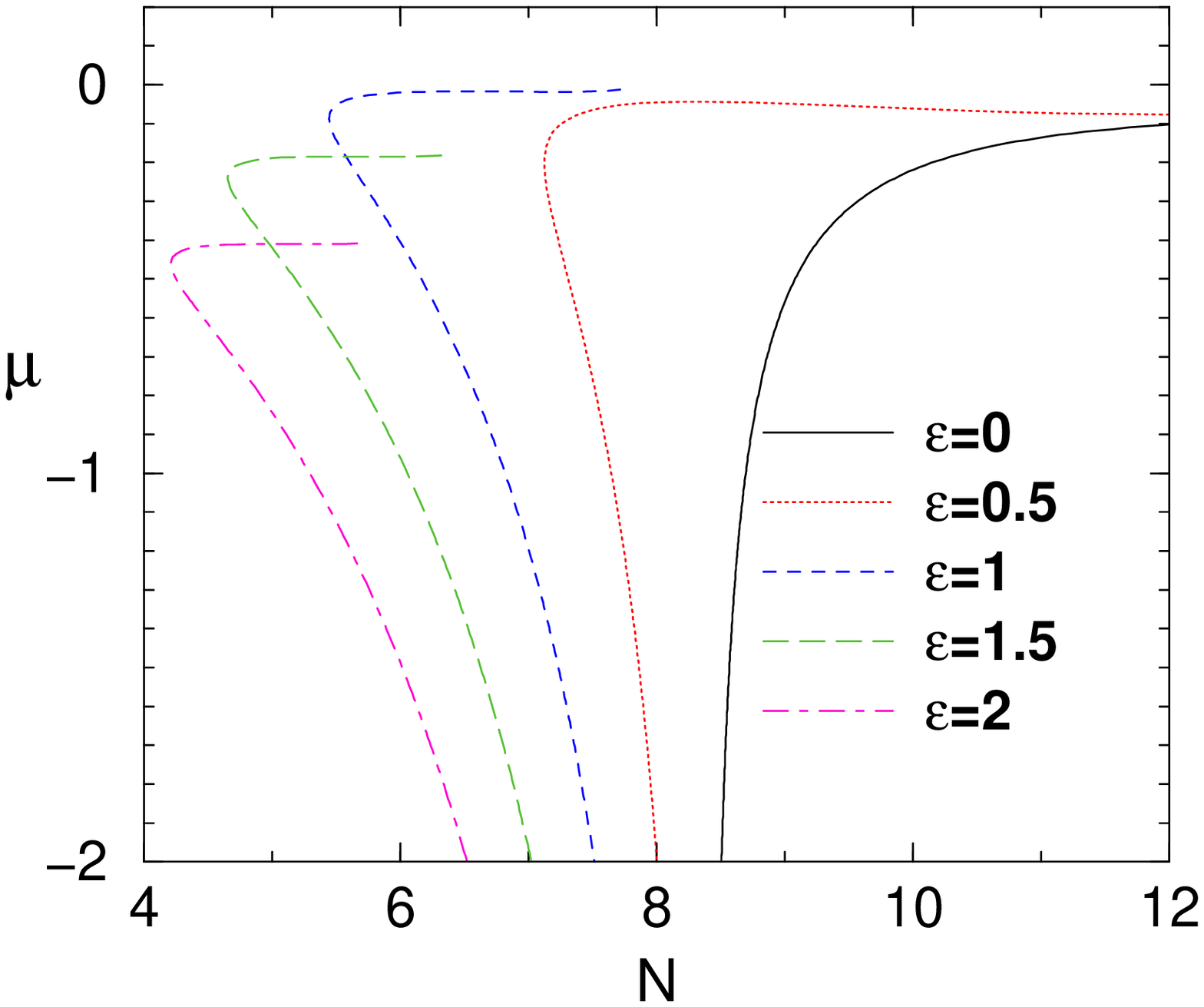}
\includegraphics[width=4.2cm,height=4.3cm,clip]{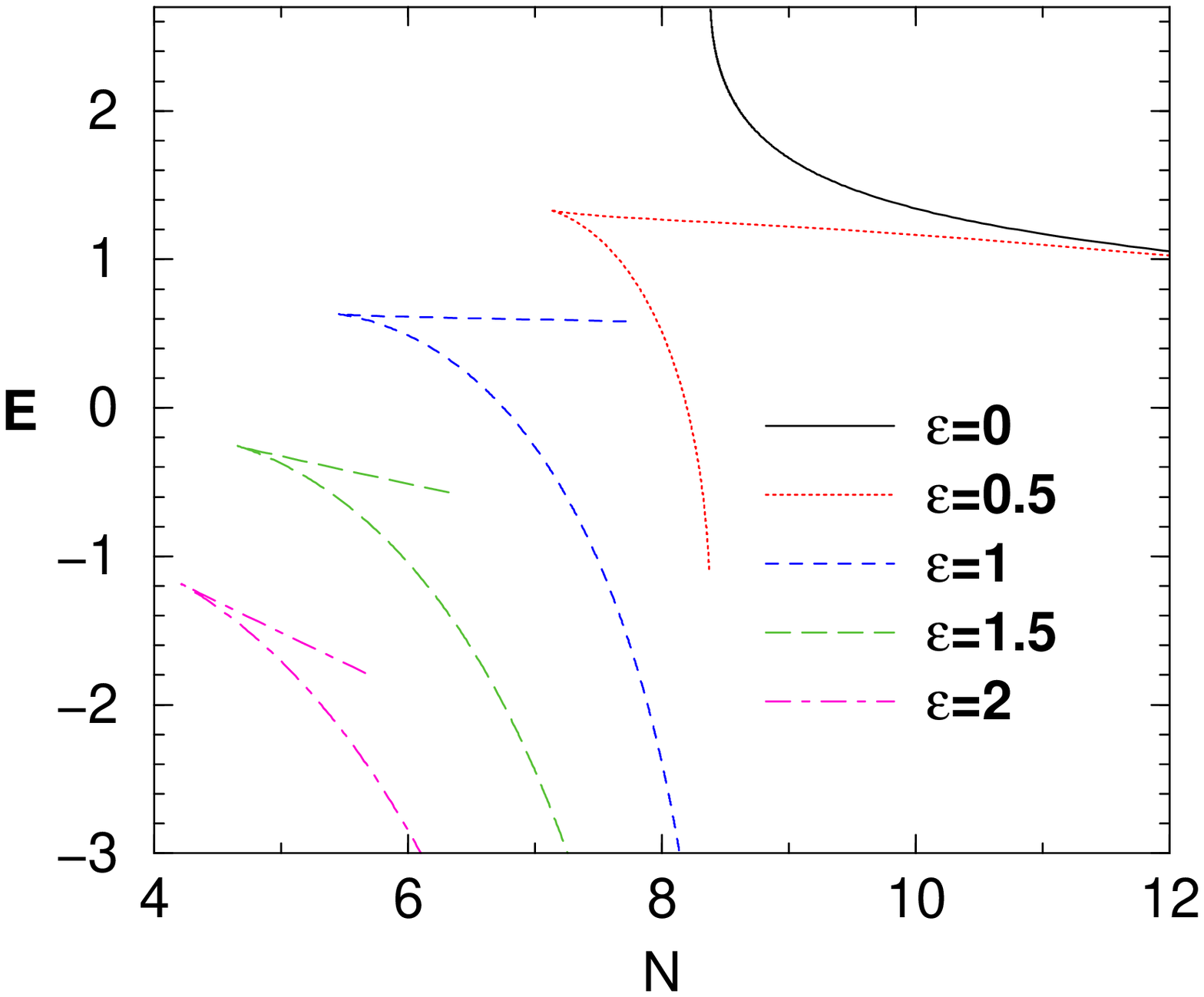}
}
\vspace{0.1cm}
\centerline{
\includegraphics[width=4.2cm,height=4.3cm,clip]{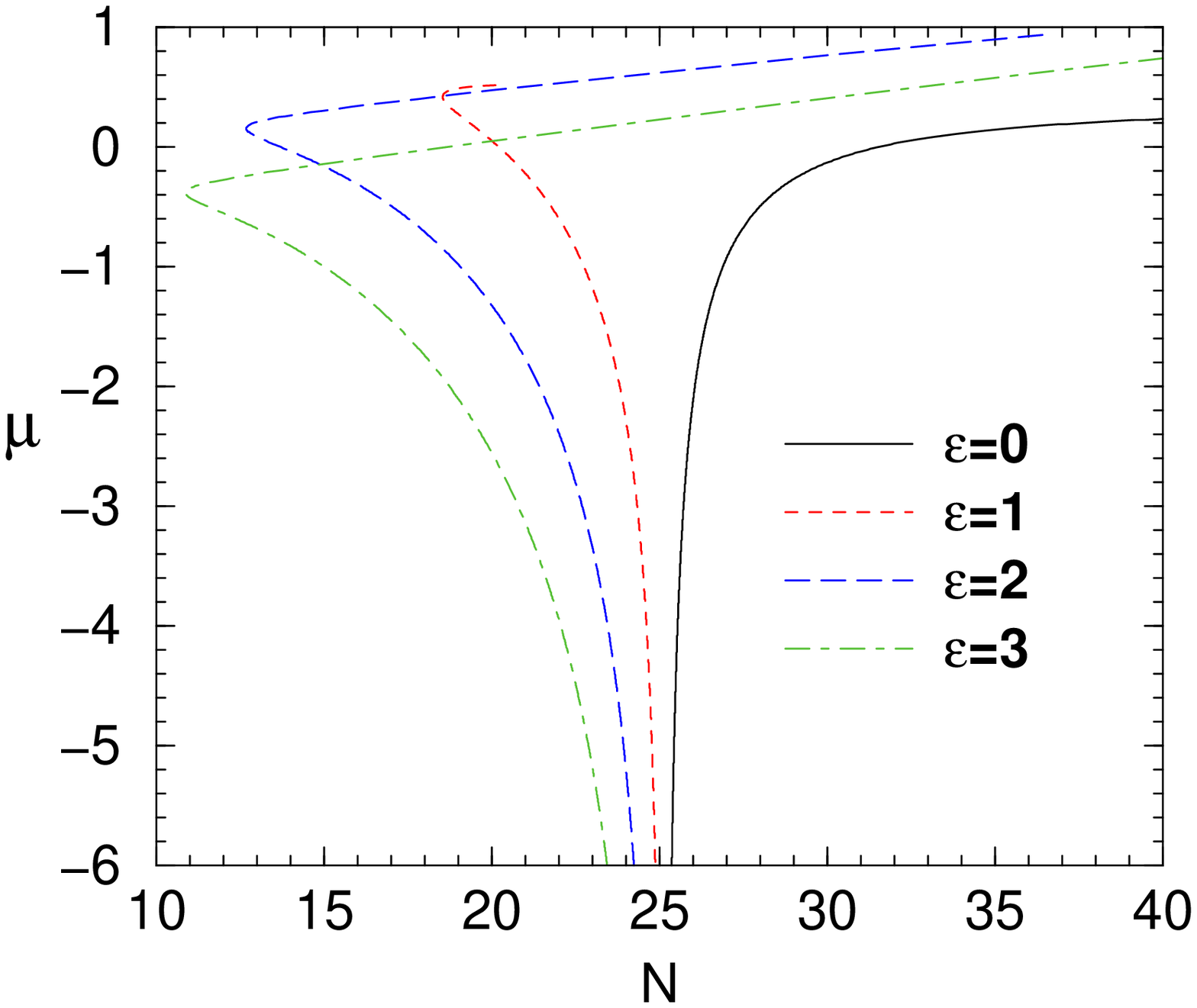}
\includegraphics[width=4.2cm,height=4.3cm,clip]{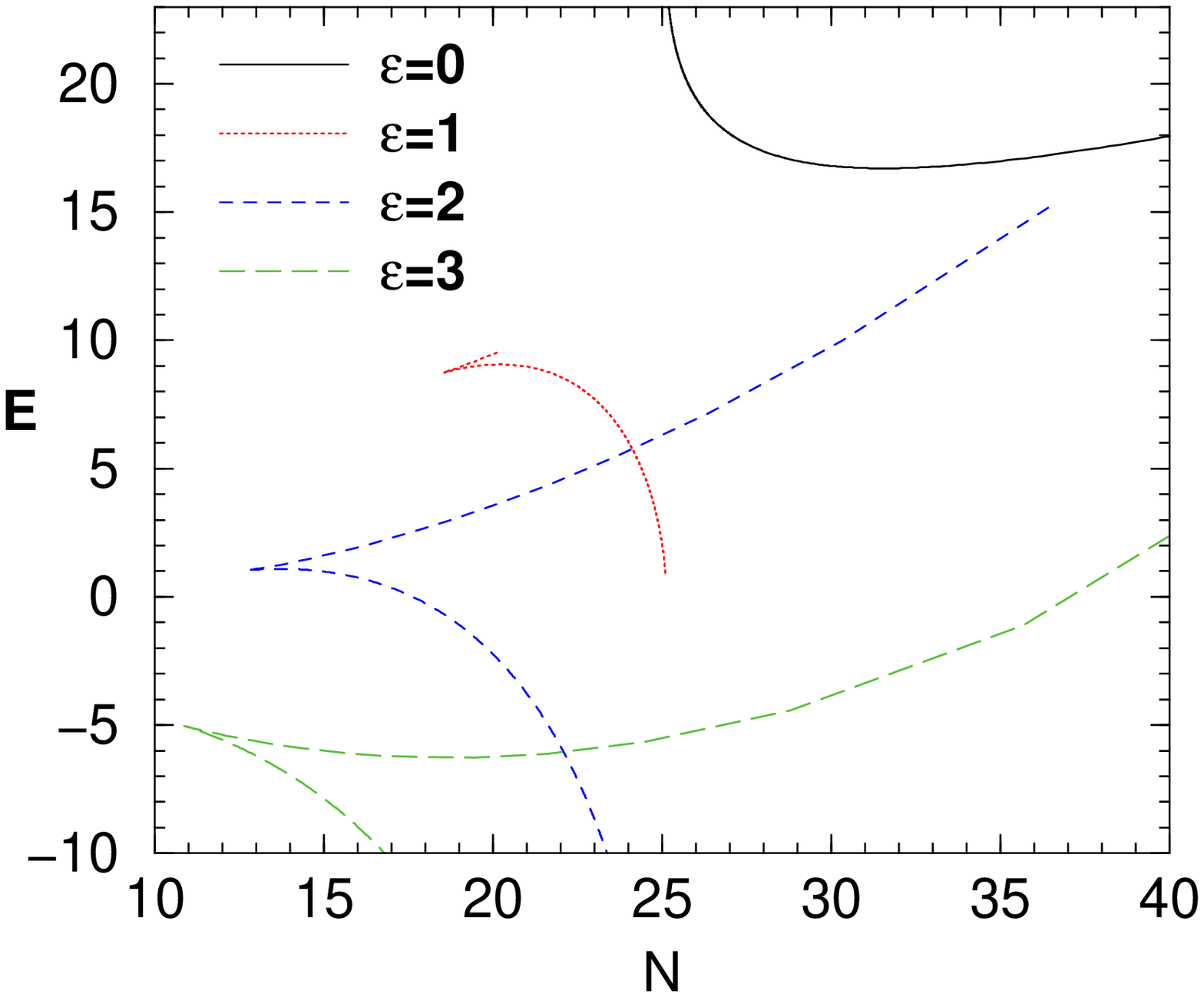}
}
\vspace*{-0.3cm}
\caption{(Color online)
Variational approach results for 2D solitons, for attractive (upper panels, with $\chi=0.5$) and repulsive (lower panels, with $\chi=-0.5$) mean nonlinearity.
The chemical potentials $\mu$ are shown in the left panels, with the total
energies $E$ shown in the right panels, as functions of $N$ for increasing
strengths of the linear OL $\varepsilon$, as indicated inside the panels.
Other parameter are fixed as $\lambda=2$, $\gamma=1.0$.
All the quantities are in dimensionless units.
}\label{Fig1}
\vskip -0.3cm
\end{figure}
In Fig.~\ref{Fig1} we  show  variational curves $N, \mu$ (left panels) obtained from the above equations for attractive (top)  and repulsive (bottom) cases. Notice the  non monotonic behavior  with the appearance of a minimum (turning point) in $N$ which gives a threshold in the number of atoms for the existence of the soliton. This is  a characteristic  of higher dimensional solitons of cubic nonlinear Schr\"odinger equations. It is associated to the so called delocalizing transition~\cite{BS04}; e.g., for values of $N$ smaller than the critical value (minimum of the curve), the solution quickly decays  in the uniform background and becomes  fully delocalized. In the right panels of Fig.~\ref{Fig1} we also show the corresponding
plots in the $E-N$ plane. Notice that, from the general definition, $\mu=\frac{d E}{dN}$, the  chemical potentials correspond to the slopes of these curves at a given $N$.
Also notice that the $E-N$ curves display cusps in correspondence of the turning points of the $\mu-N$ curves.

The stability of the soliton solutions can  be inferred  from the observed plotted results
by means of the Vakhitov-Kolokolov necessary criterion~\cite{VK}, according to which stable solution always
correspond to the branches for which $d\mu/d N <0$. We see that no solution  can be stable for both attractive and repulsive interactions in presence of only the nonlinear OL (e.g. for $\varepsilon=0$). By increasing the strength of the linear OL, however, the slope of the curves changes from positive to negative. Our analytical study therefore predicts that stable solitons in crossed linear and nonlinear OL can exist not only for  attractive interactions, a fact which is  true  also in absence of the NOL in the $y-$direction as demonstrated
in \cite{BMS04},  but also  for repulsive interactions (in absence of the NOL this last case would be impossible). It is also worth to note that the stable branches of the curve in the attractive case always lie well below the critical threshold for collapse, e.g.  $N<N_{cr}$.

In the next section we show  that  these predictions are in good agreement with the results obtained by
direct numerical integrations of the GPE.

\section{Numerical results and VA comparison}
To check the above analytical predictions we have  performed
direct numerical integrations of the 2D GPE in  Eq. (\ref{2DGPE}) with crossed linear
and nonlinear OL. The method we have used to find localized solutions  is  a
standard relaxation algorithm in imaginary time propagation with a
"back-renormalization", which allows  to fix the chemical potential $\mu$ and
determine the corresponding number of atoms $N$~\cite{marijana}.

\begin{figure}
\centerline{
\includegraphics[width=4.2cm,height=4.3cm,clip]{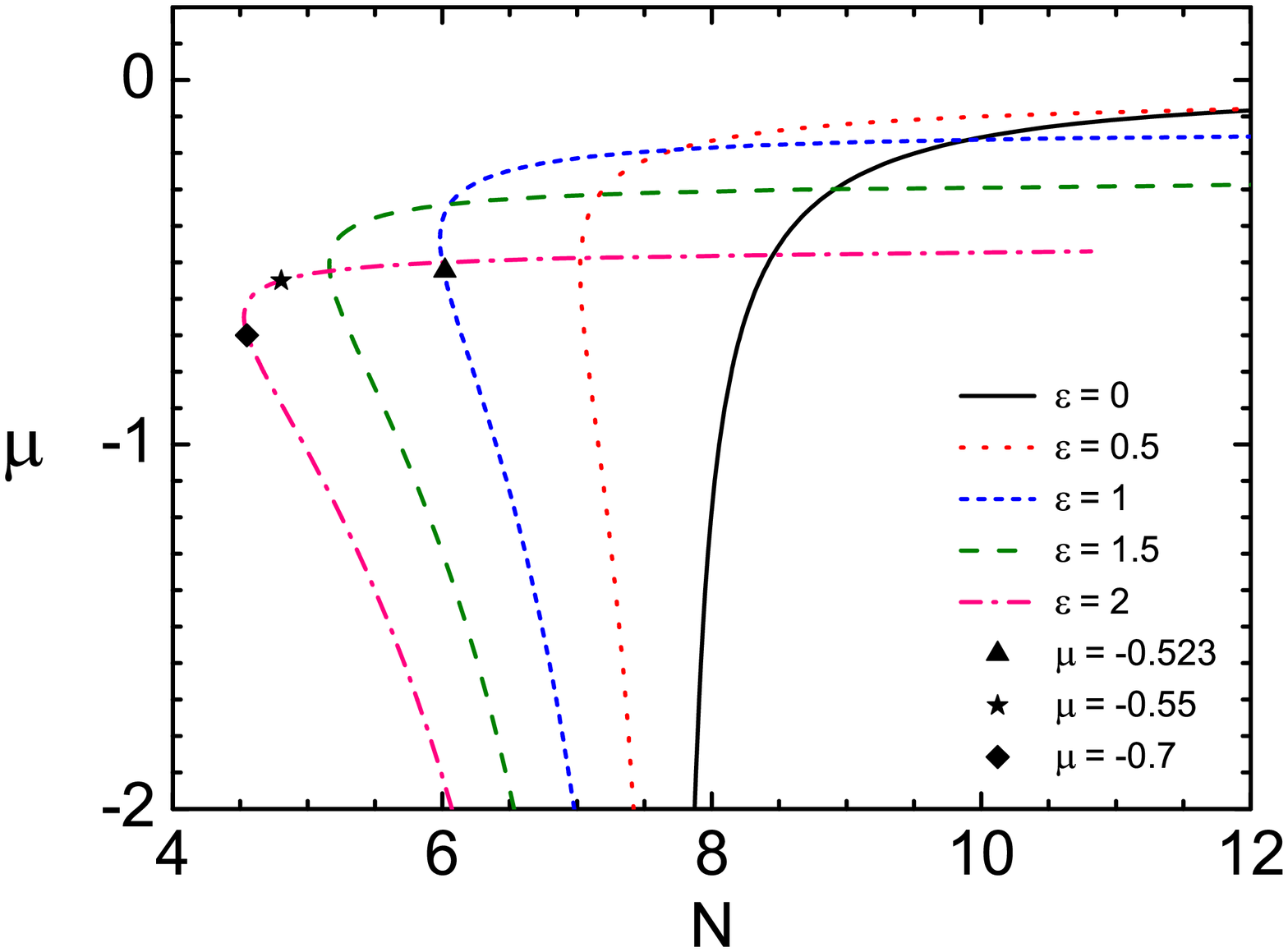}
\includegraphics[width=4.2cm,height=4.3cm,clip]{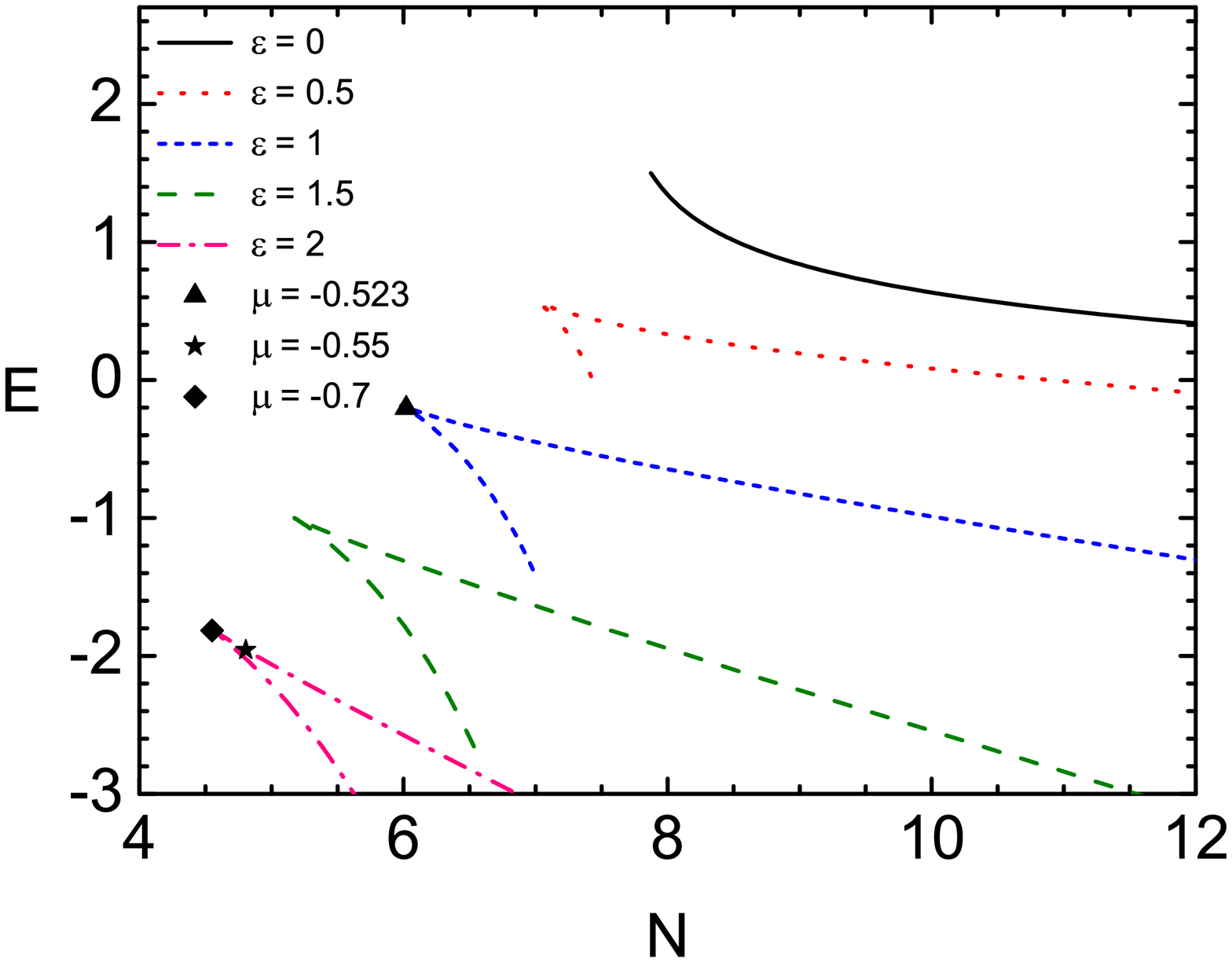}
}
\vspace{0.1cm}
\centerline{
\includegraphics[width=4.4cm,height=4.4cm,clip]{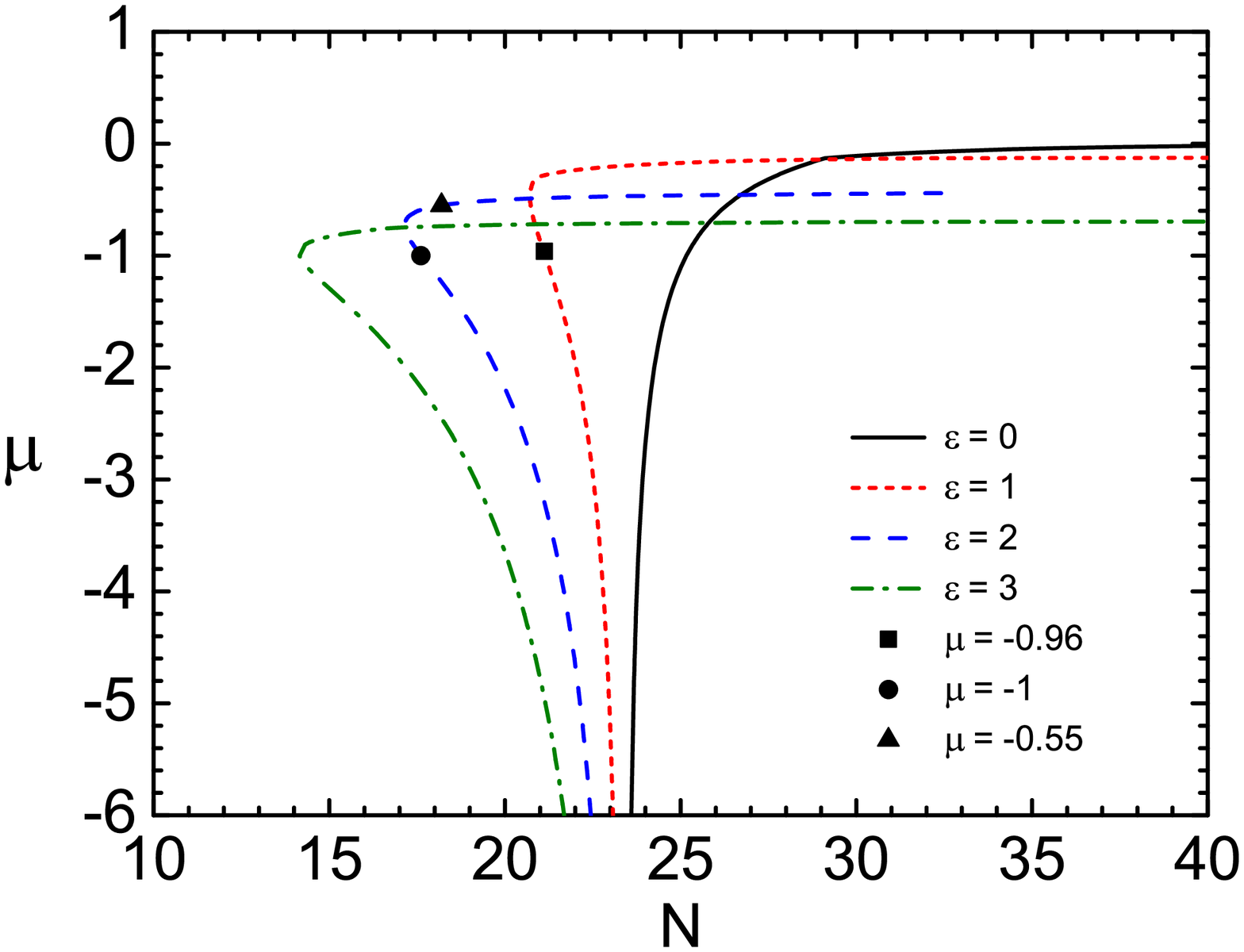}
\includegraphics[width=4.2cm,height=4.3cm,clip]{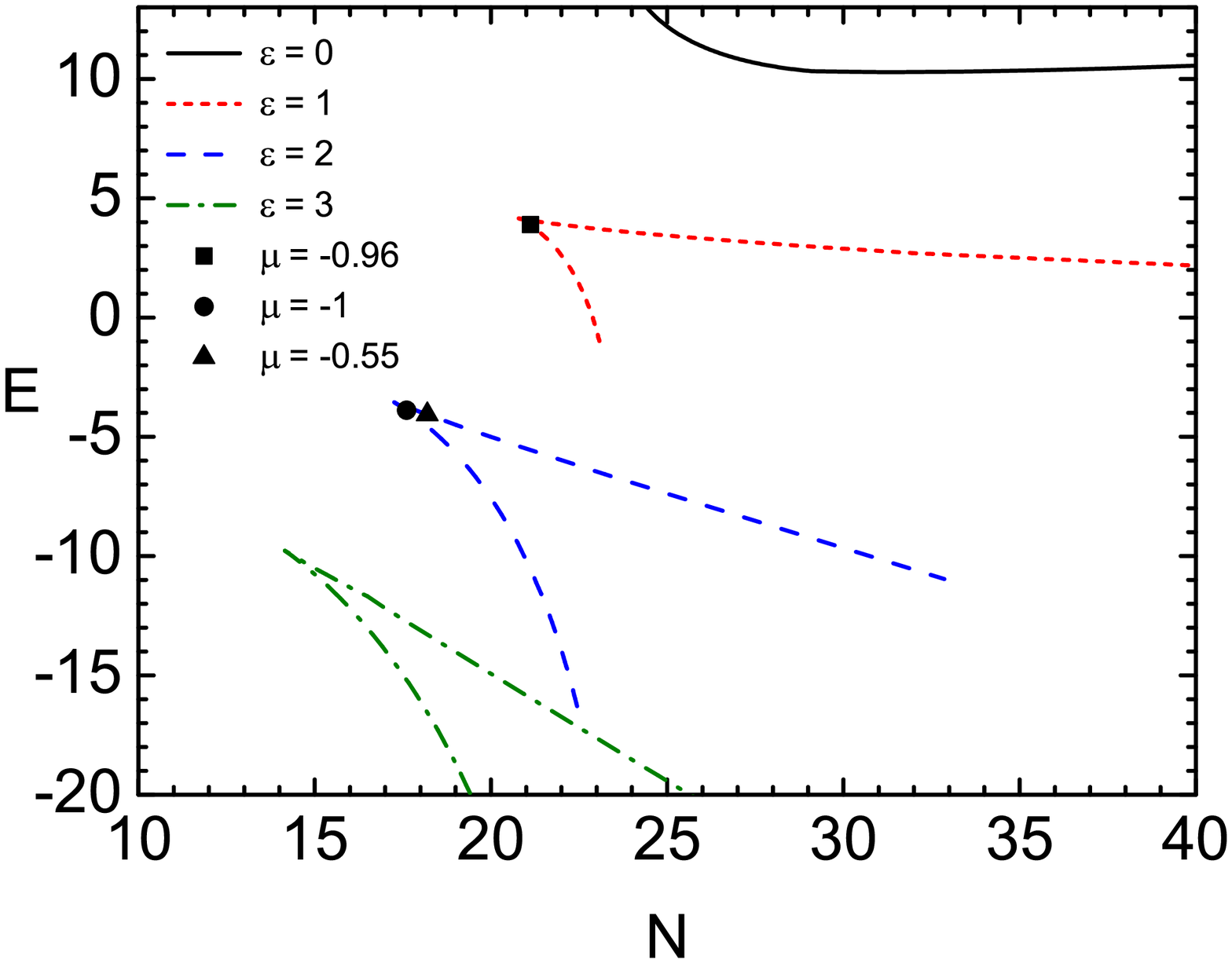}
}
\vspace*{-0.2cm}
\caption{(Color online)
Plots of 2D solitons in the $\mu-N$ (left panels) and $E-N$ (right panels) planes,
as obtained from numerical integrations of Eq.~(\ref{2DGPE}),  for attractive (upper panels, $\chi=0.5$) and repulsive (lower panels, $\chi=-0.5$) mean nonlinearity and for increasing strengths $\varepsilon$ of the linear OL (see legends inside panels). Other parameters are fixed as in Fig.~\ref{Fig1}.
Symbols on curves correspond to localized  solutions   selected for comparison with
the VA density profiles in Fig. \ref{Fig3} and for stability checks  by time propagation
in Fig. \ref{Fig4}. All the quantities are dimensionless.
}\label{Fig2}
\vskip -0.3cm
\end{figure}

\begin{figure}
\centerline{\includegraphics[width=4.5cm,height=4.5cm,clip]{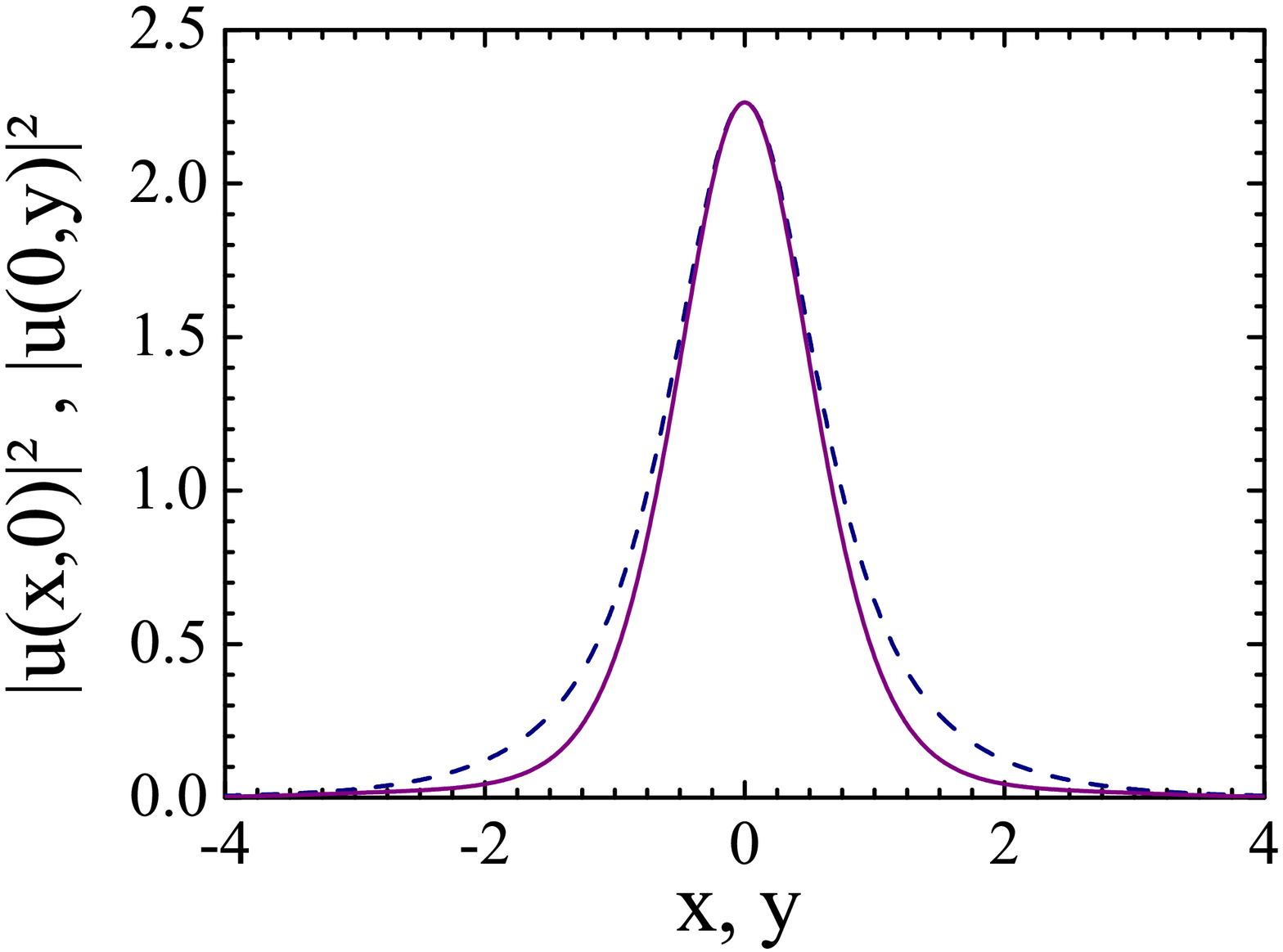}
\includegraphics[width=4.5cm,height=4.5cm,clip]{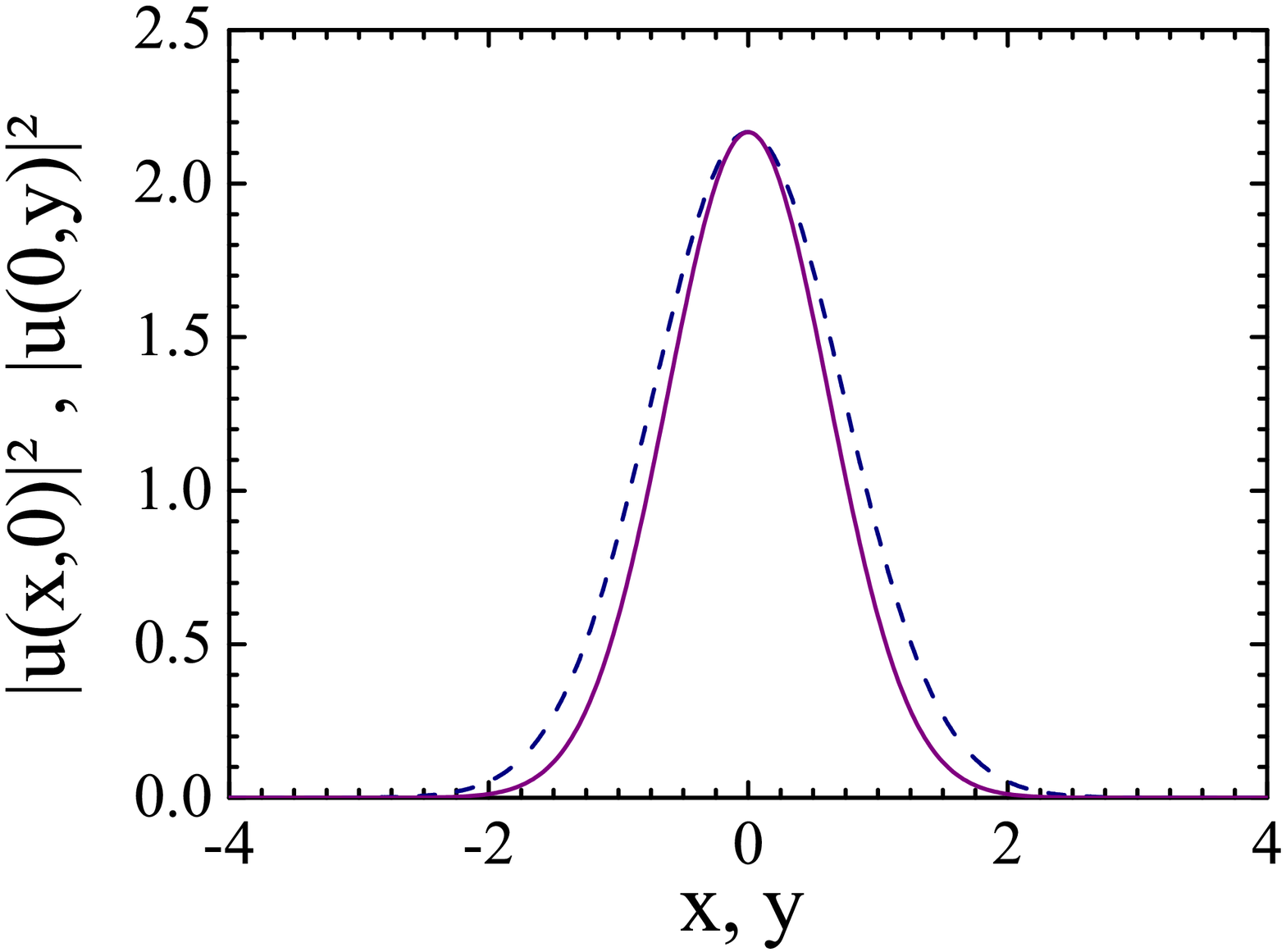}
}
\centerline{
\includegraphics[width=4.3cm,height=4.3cm,clip]{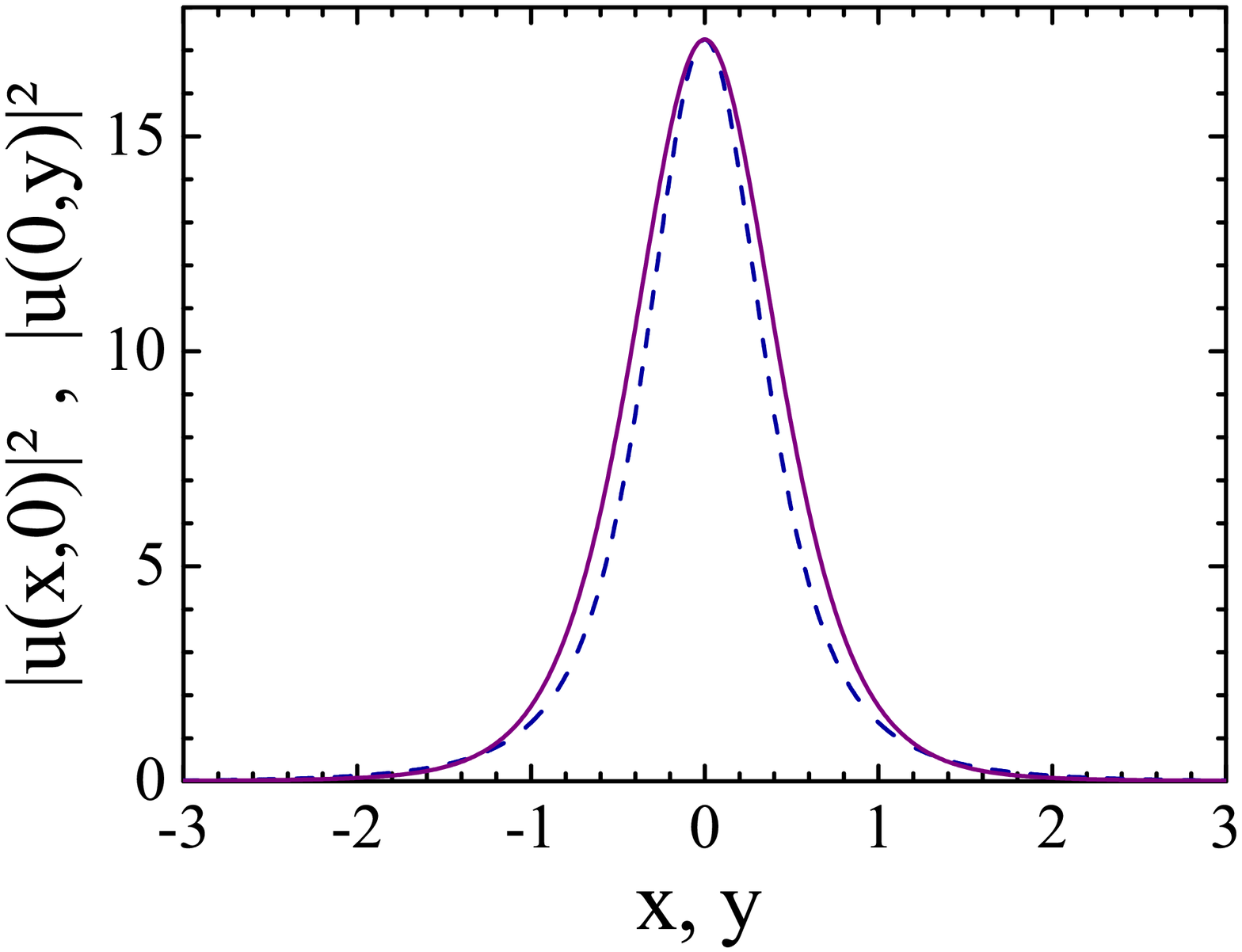}
\includegraphics[width=4.3cm,height=4.3cm,clip]{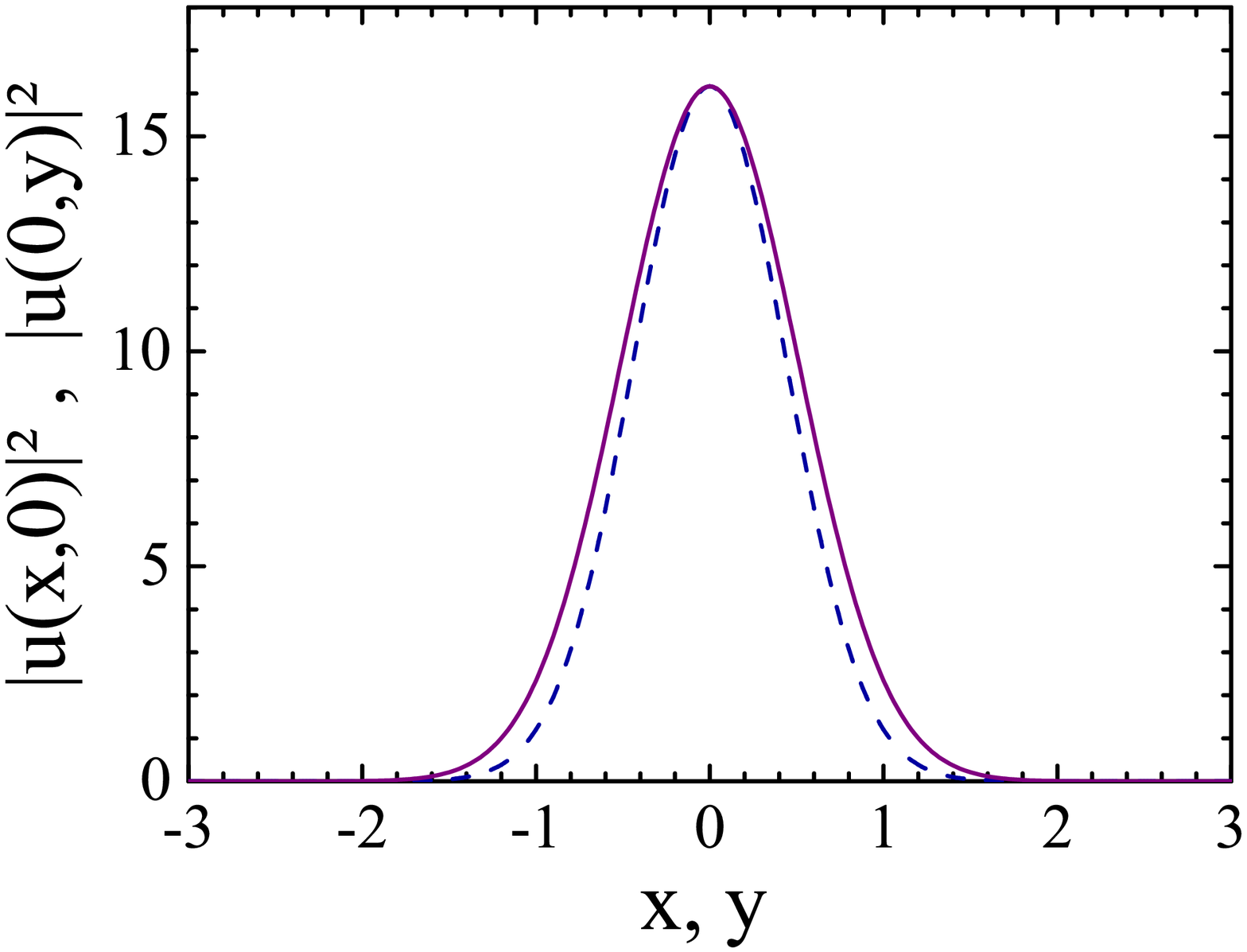}
}
\vspace*{-0.2cm}
\caption{(Color online)
Sections of the soliton density taken at $x=0$ (dashed lines) and  $y=0$
(continuous lines) for attractive $\chi=0.5$ (top panels) and repulsive $\chi=-0.5$
(bottom  panels) interactions. Left panels refer to the numerical GPE imaginary time
relaxation method calculations and correspond to the points shown on the $\varepsilon=1.0$ curves in the upper and lower left panels of Fig. \ref{Fig2},
at  $N=6.02$ ($\mu=-0.523$) and $N=21.12$ ($\mu=-0.96$), respectively. Right panels show the soliton profiles predicted by the VA for the same chemical potentials as in the corresponding left panels. In this case, for $\mu=-0.523$, we have $N=6.2$, $a=1.3$ and $b=0.928$ (attractive case); and, for $\mu=-0.96$, we have $N=22.69$, $a=1.93$ and $b=2.6$ (repulsive case).  Other parameters are fixed as in Fig.~\ref{Fig1}. All the quantities are dimensionless.
} \label{Fig3}
\vskip -0.2cm
\end{figure}

Using this method we were able to follow these solutions in the parameter space 
even in the regions where they  become  unstable (hyperbolic states), which  
permit us to trace curves in the $N-\mu$ plane and to determine the existence 
of the thresholds in the number of atoms (turning points).
Typical numerical curves for 2D solitons are depicted in Fig.~\ref{Fig2}, 
for both  attractive (top panels) and repulsive (bottom panels) interactions 
for different parameter values. Examples of soliton solutions determined with 
the relaxation method are shown in the left panels of Fig.~\ref{Fig3}.
To facilitate the comparison with the VA predictions, the same set of
parameters as the ones used in Fig.~\ref{Fig1} are considered.
We see that in the attractive case  the agreement is very good not only
qualitatively but also from a quantitative point of view.
Apart from the small  shift of the curves, the VA correctly
predicts the existence of the turning point and the change of stability of the solutions.
Notice  that the agreement improves as the strength of the LOL is increased this being
a consequence of the fact that in a deeper LOL the soliton becomes more localized and
better  described by the Gaussian  ansatz used in our variational analysis.
For the repulsive case the agreement becomes more qualitative with a larger
shift between  the numerical and the VA curves, which is unaffected by the strength of
the LOL. This discrepancy
can be ascribed to the fact that for  repulsive interactions the Gaussian ansatz used
in the VA becomes less accurate because of the tunneling of the matter into adjacent
potential wells (the condensate wavefunction cannot be localized into a single potential
well and develops satellites in adjacent wells of the OL).

In  Fig.~\ref{Fig3} we compare the sections of the density profiles of the solitons
corresponding to the points shown on the $\varepsilon=1.0$ curves in the corresponding
top and bottom left panels of Fig.~\ref{Fig2}. The right panels refer to  VA solutions
for attractive (top) and repulsive (bottom) cases,  while the corresponding profiles
obtained from numerical imaginary time integrations  of the GPE  are shown in the left
panels. We see that, for the chosen  parameters,  the agreement with our analytical
predictions is quite reasonable. Also notice that for the attractive case (positive
mean nonlinearity) the localization induced by the LOL in the $x-$direction is
stronger than the one induced by the NOL in the $y-$direction. By changing the sign
of the mean nonlinearity, however, keeping all the other parameters the same, the
opposite situation occurs.

\begin{figure}
\centerline{
\includegraphics[width=5.2cm,height=4.5cm,clip]{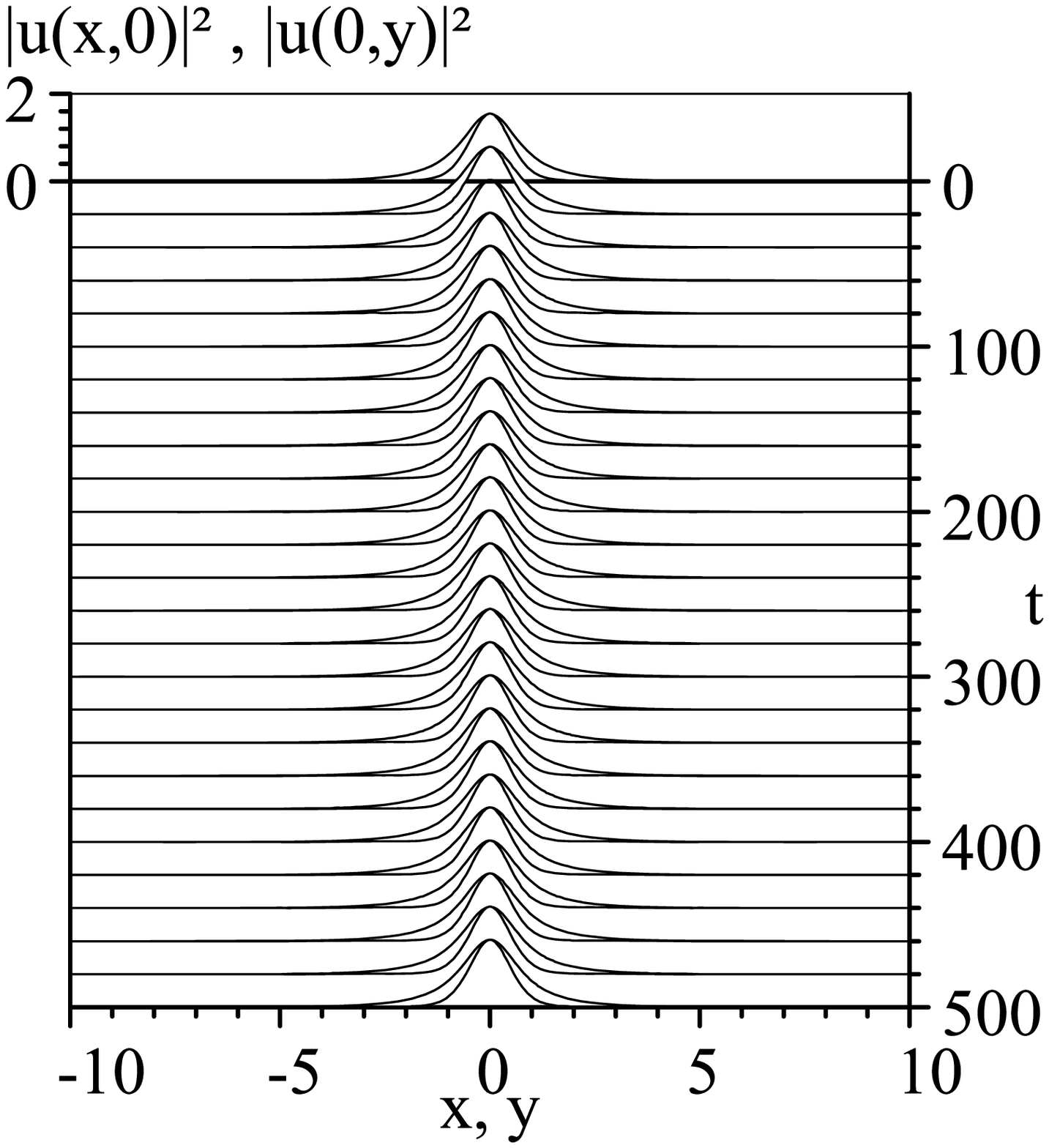}
\hspace{-1.2cm}
\includegraphics[width=5.4cm,height=4.5cm,clip]{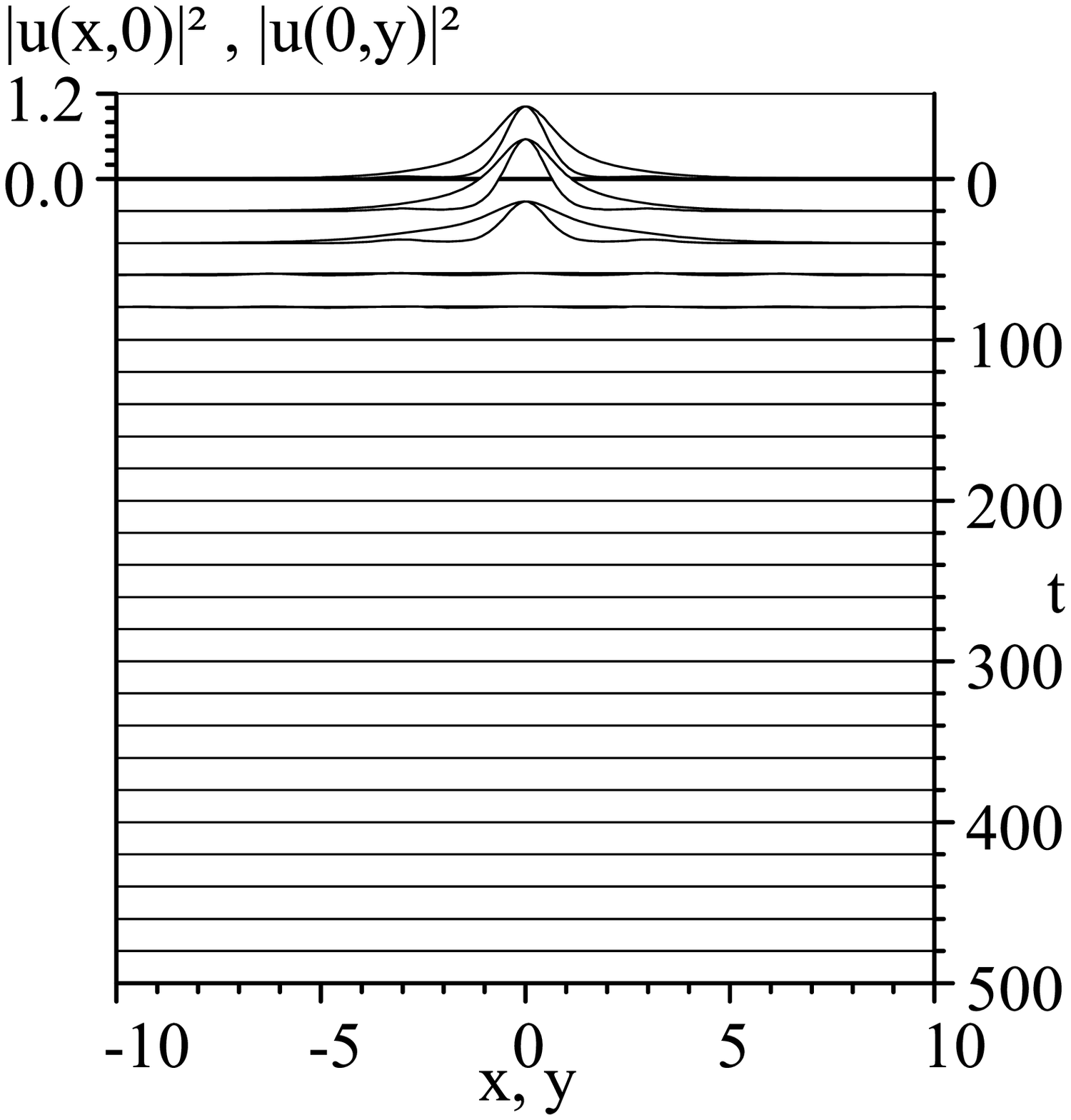}
}
\centerline{
\includegraphics[width=5.4cm,height=4.5cm,clip]{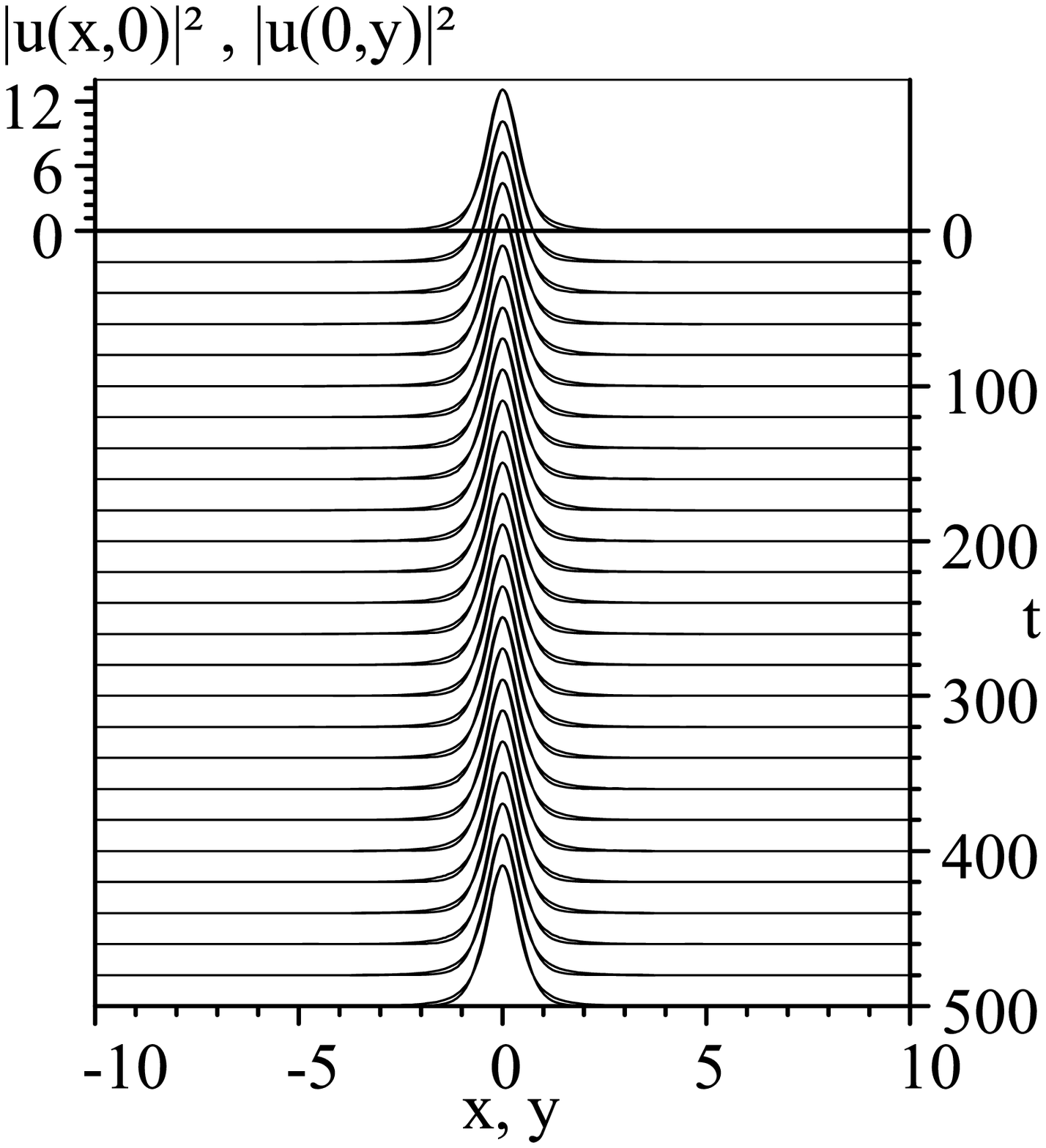}
\hspace{-1.2cm}
\includegraphics[width=5.3cm,height=4.5cm,clip]{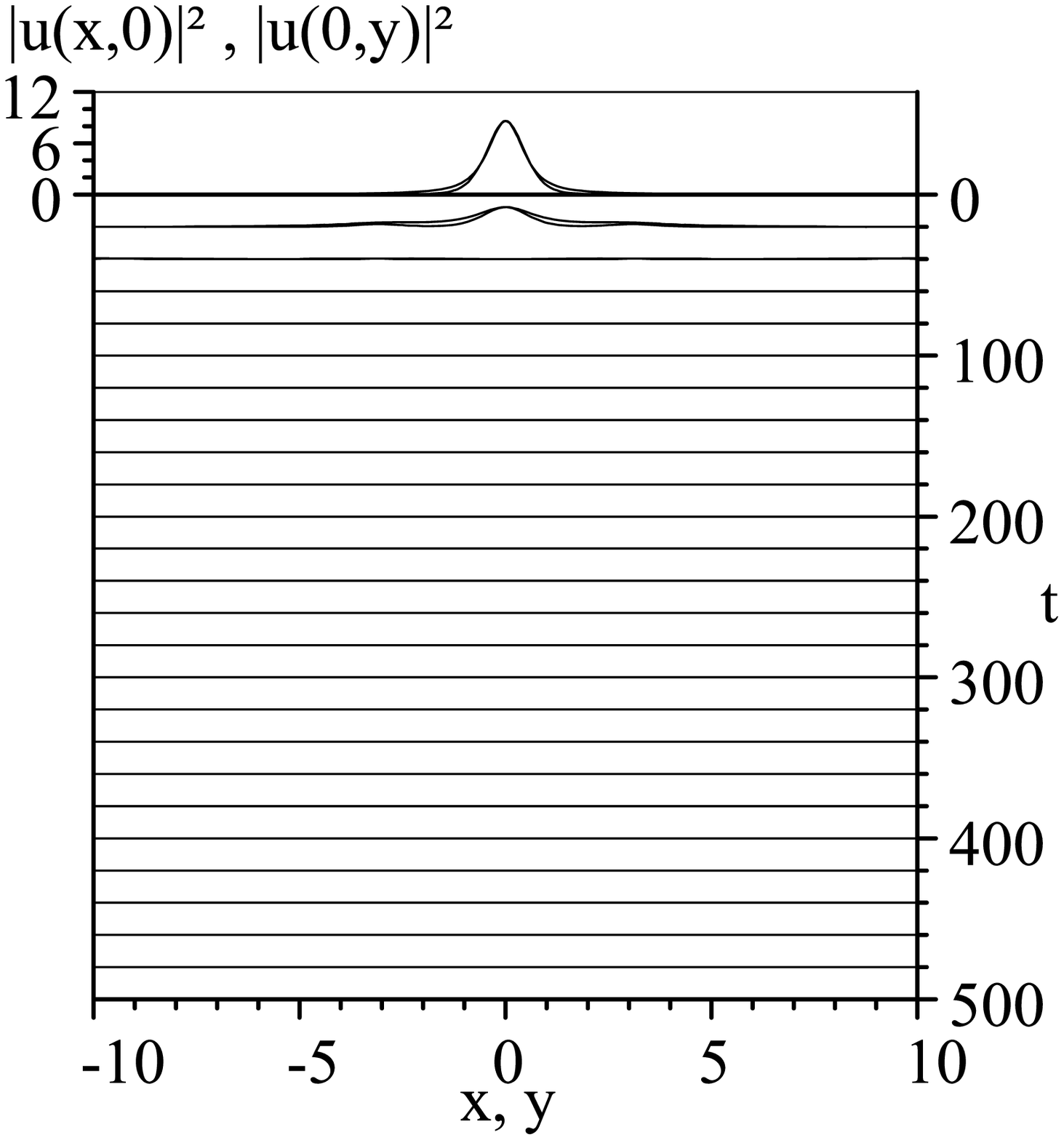}
}
\caption{Time evolution of the $x$ and $y$ sections of the soliton density
for attractive (top panels) and repulsive (bottom panels) interactions
corresponding  to the points on the stable (left panels) and unstable (right
panels) branches of the $\varepsilon=2.0$ curves in the top and bottom panels
of Fig.~\ref{Fig2}. The chemical potentials of the stable and unstable solitons
are $\mu=-0.7$ and $\mu=-0.55$ for the attractive case, with $\mu=-1.0$ (left) and
$\mu=-0.55$ for the repulsive case, respectively.
In the left side of the frames we show the density scales, with the time evolution
shown in the right side. Wider profiles correspond to y-sections.
All the quantities are dimensionless.
}
\vskip -0.3cm
\label{Fig4}
\end{figure}
The stability of these solutions has been checked  by means of a real time propagation of the GPE using a Crank-Nicolson split-method and taking as initial condition slightly perturbed solitons. In the left panels of Fig.~\ref{Fig2}, we have selected four specific soliton  solutions on branches of the curves, for $\varepsilon=$2,
with different slope close to the turning points, to check stability, e.g.  $\mu=-0.7$(diamond) and $\mu=-0.55$(star) in the attractive case(top frame); $\mu=-1.0$ (sphere) and $\mu=-0.55$(triangle) in the repulsive case (bottom frame).
In Fig.~\ref{Fig4} we show the dynamics obtained from numerical integrations of the GPE with initial conditions corresponding to these solitons with a small  perturbation added.
We see that while the  solitons on  the branch of the curves with negative slope
are stable under time evolution, the ones  on the branches with positive slope are unstable and
quickly decay into an uniform background. These results are in agreement with what is expected from our  VA analysis using the Vakhitov-Kolokolov criterion. From these studies  we conclude that it
is possible to create stable 2D BEC solitons  using a 1D linear OL  in the $x-$direction 
and a periodic modulation of the scattering length in the orthogonal directions, both for
attractive and repulsive interactions.

An estimate of the parameters for possible experimental observations of the above 2D  solitons can be made by considering, for example, the case of a repulsive BEC with  
$N \sim 10^5 $ atoms of $^{87}$Rb loaded in a crossed combined  OL consisting of a  
LOL in the $x-$direction with a period $d=2.2\mu$m, generated by a laser field of wavelength $\pi/k \approx 0.830\mu$m and strength $\Lambda=$ 1.5 $h \times$ kHz 
($h=$Planck constant), and a NOL in the $y-$direction created by an optically induced Feshbach resonance leading to a spatial variation of the scattering length of the 
form $a_s(y)=a_{s0}+a_{s1} \cos(2 \kappa y)$.
The background scattering length, $a_{s0}$, is obtained by using an uniform 2D
external magnetic field, with the value near the FR value $B_{FR}=1007$ G, which will give $a_{s0}\approx 100 a_0$, with $a_0$ denoting the Bohr radius. The  modulation part of the scattering length $a_{s1}$ is induced by illuminating the system in the $y-$direction with a laser field of wavelength  $\pi/\kappa \approx 0.784\mu$m, corresponding to the photo association wavelength of $^{87}$Rb, to shift the value of $a_{s1}$ to values $Re(a_{s1}/a_{s0}) \approx \pm (1-5)$\cite{Bauer}. Such a crossed combined  lattice  is estimated to trap  fundamental solitons (e.g solitons localized in a single potential well) containing up to few thousands of atoms.

\section{Conclusions}
By summarizing our present findings, we have shown that it is possible to
stabilize two-dimensional BEC solitons by combining a linear OL, in one direction,
with a nonlinear OL in the other orthogonal direction, where the NOL can be
obtained from a periodic modulation of the scattering length.
In particular, we have shown that in 2D crossed linear and nonlinear OLs families of 2D
solitons can exist and can be stable  for both attractive and repulsive interactions.
We have determined existent curves for chemical potentials and total energies in 
terms of the number of atoms, both by a Gaussian variational approach and  by direct numerical integrations  of the corresponding GPE.
As shown, both variational and full-numerical results have been found to be in good agreement.
The stability of the solutions has been checked in both cases, by using the
Vakhitov-Kolokolov criterion and by direct numerical real time integrations of the GPE.
In this last case, for a few set of parameters, the time evolutions of density
profiles have been  presented.
These results open the possibility to observe two-dimensional localized matter waves
in the presence of crossed linear and nonlinear optical lattices in real experiment, by
using, for example, the technique considered in Ref.~\cite{Bauer}.

\section*{Acknowledgments}
We thank Funda\c c\~ao de Amparo \`a Pesquisa do Estado de S\~ao Paulo (FAPESP)
for partial financial support. AG and LT also acknowledge partial support from
Conselho Nacional de Desenvolvimento Cient\'\i fico e Tecnol\'ogico (CNPq).
MS acknowledges FAPESP and the MIUR (PRIN-2008 initiative) for partial
financial support. FKA is supported by a Marie Curie IIF under the grant
PIIF-GA-2009-236099(NOMATOS).
HFLdL also wish to thank the Department of Physics ``E.R. Caianiello" for the
hospitality for a three month stay at the University of Salerno as external
PhD student.

\end{document}